\documentclass{article}
\usepackage[top=2cm,bottom=2cm,left=2cm,right=2cm]{geometry}
\usepackage{graphicx}
\usepackage{epstopdf, epsfig}
\usepackage[colorlinks = true,
            linkcolor = blue,
            urlcolor  = blue,
            citecolor = blue,
            anchorcolor = blue]{hyperref}

\newcommand\redsout{\bgroup\markoverwith{\textcolor{red}{\rule[0.5ex]{2pt}{0.4pt}}}\ULon}

\usepackage[dvipsnames]{xcolor}
\usepackage{amsmath}
\usepackage{authblk}
\usepackage{bm}
\usepackage[normalem]{ ulem }
\usepackage{soul}
\usepackage{textcomp} 
\usepackage{float}
\usepackage{color}
\usepackage{tikz}
\usetikzlibrary{arrows.meta}
\usetikzlibrary{calc}
\usetikzlibrary{math}
\usetikzlibrary{angles, quotes}
\usepackage{caption}
\usepackage{array,multirow,makecell}
\usepackage{enumitem}
\usepackage{wrapfig}
\usepackage{titlesec}
\usepackage{lmodern}
\usepackage{cases}
\usepackage{siunitx}
\usepackage{csquotes}
\usepackage{empheq}
\usepackage{mathrsfs} 
\usepackage{subcaption}

\newcommand{\pd}{\right)}
\newcommand{\pg}{\left(}
\newcommand{\md}{\right>}
\newcommand{\mg}{\left<}
\newcommand{\cd}{\right]}
\newcommand{\cg}{\left[}

\newcommand{\son}{c}
\newcommand{\dens}{\rho_0}

\newcommand{\refr}{\mathcal{R}}
\newcommand{\vit}{U}

\newcommand{\surf}{\mathcal{S}}

\newcommand{\sinf}{\mathcal{S_\infty}}
\newcommand{\ninf}{\mathbf{n_\infty}}
\newcommand{\rinf}{r_{\infty}}
\newcommand{\vol}{\mathcal{V}}

\newcommand{\vun}{\mathbf{v_1}}

\newcommand{\pun}{p_1}

\newcommand{\dun}{\rho_1}

\newcommand{\force}{\mathbf{F}_{\text{rad}}}

\newcommand{\ide}{\Bar{\Bar{I}}}

\renewcommand{\d}{\mathrm{d}}
\renewcommand{\div}{\mathrm{\nabla\cdot}}

\title{Acoustic dipole surfing on its own acoustic field: toward acoustic quantum analogues.}
\author[1]{J.-P. Martischang}
\author[1]{A. Roux}
\author[,1,2]{M. Baudoin\thanks{Email address for correspondence: michael.baudoin@univ-lille.fr}}
\affil[1]{Univ. Lille, CNRS, Centrale  Lille, Univ. Polytechnique Hauts-de-France, UMR 8520, IEMN, F59000 Lille, 
France}
\affil[2]{Institut Universitaire de France, 1 rue Descartes, 75005 Paris}

\newcommand{\paper}{[J. Fluid Mech., 952: A22 (2022)]}

\begin{document}

\maketitle

\begin{abstract}
In a recent paper \paper, Roux et al. demonstrated that a translating monopolar acoustic source is subjected to a self-induced radiation force opposite to its motion. This force results from a symmetry breaking of the emitted wave induced by Doppler effect.  In the present work, we show that for a dipolar source, the self-induced radiation force can be aligned with the velocity perturbation, hence amplifying it. This work suggests the possibility of a dipolar acoustic source surfing on its own acoustic wave, hence paving the way towards acoustic quantum analogues.
\end{abstract}

{\bfseries \itshape{Keywords :}}
acoustic radiation force - dipole - quantum analogues

\section{Introduction}

In their seminal experiments, \cite{nature_couder_2005} unveiled a classical hydrodynamic system exhibiting a wave-particle duality. This system is made of a self-propelling drop driven by a resonant interaction with its own wavefield. This wavefield is created by a drop bouncing (without merging) at the surface of a bath excited just below the Faraday threshold (\cite{ptrsl_faraday_2022,jfm_kumar_1994}) and giving birth to a localized wave intrinsically linked to its dual particle. With this system, researchers were able to reproduce a wealth of behaviors classically associated with the quantum realm such as: tunneling (\cite{prl_eddi_2009,pre_hubert_2017,prf_nachbin_2017},\cite{pre_tadrist_2020}), Landau levels (\cite{pnas_fort_2010,jfm_harris_2014,jfm_oza_2014}), quantum coral (\cite{pre_harris_2013,pre_gilet_2014,pre_gilet_2016,np_saenz_2018,c_christea_2018}) or Friedel oscillation (\cite{sa_saenz_2020}). An exhaustive list of the quantum analogies explored with these self-propelling drops can be found in the recent review by \cite{rpp_bush_2020}.

While this system has considerably contributed to the development of the field of classical quantum analogues, it also has some inherent limitations. For example, the possibilities and limits of these hydrodynamic quantum analogues to reproduce the iconic double-slits experiment have been thoroughly explored by \cite{pre_andersen_2015,njp_dubertrand_2016,jfm_faria_2017,jfm_pucci_2018,prf_rode_2019,pre_ellegaard_2020} after the initial experimentation by \cite{prl_couder_2006}. More generally, (i) hydrodynamic quantum analogues are only 2D by nature, while quantum mechanics is inherently 3D. (ii) The influence of the hydrodynamic pilot wave through memory effects is limited by viscous damping. Third, only monopolar bath vibrations can be created by the droplet bouncing, excluding the exploration of other modes. Finally, the system requires a periodic disconnection between the particle and its wave in an additional spatial dimension, that cannot exist in a 3D system.

Acoustic quantum analogues could provide an elegant solution to some of these limitations. Indeed, it is known since the early work by \cite{pm_rayleigh_1905}, \cite{jpr_brillouin_1925} and Langevin (work reported later on by \cite{ra_biquard_1932a,ra_biquard_1932b}), that an acoustic scatterer excited by an acoustic field is subjected to a nonlinear force called the acoustic radiation force. This force results from intrinsic nonlinearities of Navier-Stokes equations and another nonlinearity resulting from the integration of the stress exerted by the acoustic wave on a vibrating surface (\cite{jpr_brillouin_1925}). The acoustic radiation force can be used to remotely trap and manipulate objects (see e.g. the review by \cite{arfm_baudoin_2020}). This has led to the development of standing-wave based and vortex based selective acoustical tweezers (\cite{nc_marzo_2015,baresch2016observation,prap_riaud_2017,sa_baudoin_2019,baudoin2020naturecell}), with a renewed interest in the last century to manipulate millimeter down to micrometer scale objects both in vitro and in vivo (\cite{nrmp_rufo_2022}). This interest in acoustic manipulation has also led to some recent theoretical developments to compute the radiation force (\cite{jasa_silva_2011,baresch2013three,sapozhnikov2013radiation,jasa_gong_2021}) and torque (\cite{silva2012radiation,jasa_gong_2020}) exerted by an arbitrary acoustic field on a spherical particle.

Yet, the emergence of an acoustic wave-particle duality requires that the acoustic source is moved by its own acoustic field, not by an incident field. Also ideally, the direction of motion of the source should not be set a priori by an anisotropy of the radiated field. For symmetry reasons one can nevertheless easily figure out that an isotropic acoustic field cannot give birth to a directional self-induced radiation force. Recently, \cite{jfm_roux_2022} demonstrated theoretically that adding a slight translation perturbation to a monopolar source induces a symmetry breaking due to Doppler effect and hence a directional acoustic radiation force. But in this case, the force was opposite to the velocity perturbation, hence leading to a stable configuration. Here we compute the self-induced radiation force exerted on a dipolar source. We show that depending on the orientation of the dipole compared to the velocity perturbation, the radiation force can be oriented in the same direction as the velocity perturbation, hence amplifying it. This unstable situation paves the way towards acoustic sources driven by their own acoustic field, a situation analogous to the one at the origin of hydrodynamic quantum analogues. Note that this situation is reminiscent of some recent discoveries of hydrodynamic surfers driven by they own capillary and gravity waves (\cite{arxiv_ho_2021,prf_benham_2022}), though in these cases, the direction of motion is set a priori by the orientation of the source.

\section{Dipolar source}

The calculation of the acoustic radiation force for a dipolar source follows the same essential steps as the ones used previously to compute it for a monopolar source (\cite{jfm_roux_2022}), but with an additional degree of complexity induced by the vectorial nature of the source and the orientation between the source and the translation direction.

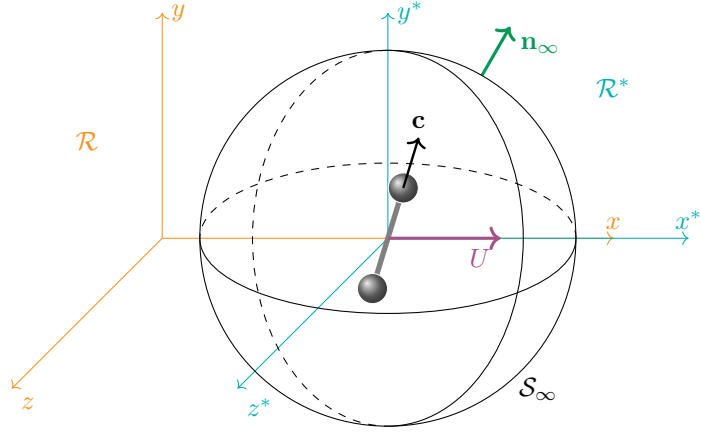
\begin{figure}
	\centering
	\begin{tikzpicture}[scale=1]
\draw[->,color=BurntOrange] (0,0)--++(6,0) node [above] {$x$};
\draw[->,color=BurntOrange] (0,0)--++(0,3) node [right] {$y$};
\draw[->,color=BurntOrange] (0,0)--++(-2,-2) node [below right] {$z$};
\node[color=BurntOrange] at (-1,1.3) []{$\refr$};

\draw[->,color=TealBlue] (3,0)--++(4,0) node [above] {$x^*$};
\draw[->,color=TealBlue] (3,0)--++(0,3) node [right] {$y^*$};
\draw[->,color=TealBlue] (3,0)--++(-2,-2) node [below right] {$z^*$};
\node[color=TealBlue] at (6,2) []{$\refr^*$};

\coordinate (O) at (3,0) ;
\coordinate (A) at ($(O)+(\angdip:0.7)$) ;
\coordinate (B) at ($(O)-(\angdip:0.7)$) ;
\draw[color=gray, line width=2pt] (A)--(B) ;
\draw [color=white,ball color=gray,smooth] (A) circle (0.2);
\draw [color=white,ball color=gray,smooth] (B) circle (0.2);
\draw[->, line width=0.9] (A)--($(A)+(\angdip:0.7)$) node[above]{$\mathbf{c}$};

\draw [->,line width=1.2 pt,color=DarkOrchid] (3,0) --++ (1.5,0) node [below left] {$U$};
\coordinate (N) at ($(O)+(60:2.5)$);
\draw [->,line width=1.2 pt,color=ForestGreen] (N) -- ($(N)+(60:0.75)$) node [below right] {$\ninf$};

\draw [-] (3,0) circle (2.5);
\draw [-] (5.5,0) arc (0:-180:2.5 and 1) ;
\draw [dashed] (5.5,0) arc (0:180:2.5 and 1) ;
\draw [dashed] (3,2.5) arc (90:270:1.8 and 2.5) ;
\draw [-] (3,2.5) arc (90:-90:1.8 and 2.5) ;
\node[] at (5,-2) []{$\surf_\infty$};
	\end{tikzpicture}
	\caption{Sketch of the problem. $\mathcal{R}=(O,(x,y,z),t)$ is a Galilean reference frame. $\mathbf{U}$ corresponds to the velocity translation of the dipolar source and $\mathcal{R}^*=(O,(x,^*y^*,z^*),t)$	to a reference frame linked to the source. The surface $\sinf$ is a spherical surface centered on the dipole, moving with it at the velocity $\mathbf{\vit }$ in $\refr$ and with a radius  $r_\infty$ much larger than all other characteristic length of the problem.}
	\label{fig:schema} 
\end{figure}

The first step is to model the dipolar source. In acoustics, a fixed punctual dipolar source can be seen as a punctual volumetric source of force: $$\mathbf{f_v}(\mathbf{r},t) = \mathbf{f}(t)\cdot\delta(x)\cdot\delta(y)\cdot\delta(z),$$ with $t$ and $(x,y,z)$ the time and Cartesian coordinates associated with a Galilean reference frame $\mathcal{R}=(O,(x,y,z),t)$, $\delta$ the Dirac delta function, $\mathbf{r}=(x,y,z)$ the position vector and $\mathbf{f}$ the vector function specifying the time dependence of the the volumetric force. If we consider a small translation perturbation of this source under the form of a constant speed $\mathbf{U}=M c \; \mathbf{x}$ along the axis $Ox$ (Fig.\ref{fig:schema}), with $M \ll 1$ the Mach number and $c$ the sound speed, the translating source can be modeled by the source term:
    \begin{equation}
        \mathbf{f_v}(\mathbf{r},t) = \mathbf{f}(t)\cdot\delta(x-Mct)\cdot\delta(y)\cdot\delta(z).
    \end{equation}
For the sake of generality, the direction of the dipolar source is supposed to be arbitrary, leading to the general expression: 
    \begin{equation}
        \mathbf{f}(t) = F  \sin(\omega t) 
        \begin{pmatrix}
            c_{x}\\
            c_{y}\\
            c_{z}\\
        \end{pmatrix}
        \label{eq:f}
    \end{equation}
    with $\mathbf{c}=\begin{pmatrix}
        c_x\\c_y\\c_z
    \end{pmatrix}$ a unit vector $(\|\mathbf{c}\|=1)$, $F$ the force amplitude, $\omega$ the angular frequency and $\|\mathbf{.}\|$ the Euclidean norm operator.
    
\section{Wavefield radiated by a translating dipolar source}

    \subsection{Wave equation}
The wavefield radiated by a translating dipolar source can be computed by solving the linearized Euler equations (mass and momentum balances):
        \begin{subequations}
        \begin{align}
            & \frac{\partial\rho_1}{\partial t}+\rho_0\cdot\div(\mathbf{v_1}) = 0,\label{eq:continuity}\\
            & \rho_0\frac{\partial\mathbf{v_1}}{\partial t} +\nabla p_1-= \mathbf{f}(t)\cdot\delta(x-Mct)\cdot\delta(y)\cdot\delta(z), \label{eq:linmomentum}
        \end{align}
        \end{subequations}
with $\rho_1$, $p_1$ and $\mathbf{v_1}$ the first order perturbations in density, pressure and velocity respectively, and $\rho_0$ the density at rest. In addition, the linearized equation of state for an inviscid fluid reads: $p_1/\rho_1 = c^2$ with $c$ the sound speed.

(i) Subtracting $c$ times the gradient of eq. (\ref{eq:continuity}) to the time derivative of eq. (\ref{eq:linmomentum}), (ii) taking into account the equation of state $p_1/\rho_1 = c^2$ and the irrotational nature of the acoustic wavefield leading to $\nabla(\div\mathbf{v_1})=\Delta \mathbf{v_1}$, (iii) introducing the acoustic displacement $\mathbf{u_1}$ such that $\mathbf{v_1} = \frac{\partial\mathbf{u_1}}{\partial t}$ and (iv) integrating the resulting equation over time leads to the classic wave equation for the displacement with the dipolar source on the rhs:
        \begin{equation}
            \frac{\partial^2\mathbf{u_1}}{\partial t^2} - c^2\Delta\mathbf{u_1} = \frac{1}{\rho_0} \mathbf{f}(t)\cdot\delta(x-Mct)\cdot\delta(y)\cdot\delta(z)\;.
            \label{eq:wave}
        \end{equation}
        
    \subsection{Lorentz transformation}
Using the Lorentz invariance of the wave equation, this equation can be rewritten in a reference frame wherein the source is fixed with the following change of variables:
        \begin{subnumcases}{}
            x' = \gamma(x-Mct)\\
            y' = y\\
            z' = z\\
            ct' = \gamma(ct-Mx)
        \end{subnumcases}
        where $\gamma=\frac{1}{\sqrt{1-M^2}}$. This Lorentz transform allows to rewrite equation (\ref{eq:wave}) into: 
        \begin{equation}
            \frac{\partial^2\mathbf{u_1}}{\partial t'^2} - c^2\Delta'\mathbf{u_1} = \frac{1}{\rho_0}\mathbf{f}\left(\frac{\gamma}{c}\left(ct'+Mx'\right)\right)\delta\left(\frac{x'}{\gamma}\right)\delta(y')\delta(z').
        \end{equation}
        We can see that for any $x'\neq0$, the rhs term is null, which means that we can consider it only when $x'=0$. If we combine this with the fact that $\delta\left(\frac{x'}{\gamma}\right)=\gamma\delta(x')$, we can write:
        \begin{equation}
            \frac{\partial^2\mathbf{u_1}}{\partial t'^2} - c^2\Delta'\mathbf{u_1} = \frac{\gamma}{\rho_0}\mathbf{f}(\gamma t') \delta(x')\delta(y')\delta(z').
        \end{equation}

        This equation can be further simplified using the second transform:
        \begin{subnumcases}{}
            x'' = \gamma x'\\
            y'' = \gamma y'\\
            z'' = \gamma z'\\
            ct'' = \gamma ct'
        \end{subnumcases}
leading to the final equation:
        \begin{equation}
            \Delta''\mathbf{u_1}-\frac{1}{c^2}\frac{\partial^2\mathbf{u_1}}{\partial t''^2} = -\frac{\gamma^2}{c^2\rho_0}\mathbf{f}(t'') \delta(x'')\delta(y'')\delta(z'').
            \label{eq:second_transform}
        \end{equation}
        
\subsection{Solutions of the equation}
The solution of the wave equation for a fixed punctual source is well known (see e.g. \cite{mgbc_morse_1968}):
        \begin{equation}
            \mathbf{u_1} =\displaystyle \frac{\gamma^2}{c^2\rho_0}\frac{\mathbf{f}\left(t''\pm\frac{r''}{c}\right)}{4\pi r''},
        \end{equation}
        which becomes with the initial variables (see the details in \cite{jfm_roux_2022}):
        \begin{equation}
        \boxed{
            \mathbf{u_1} = \frac{\mathbf{f}\left(t-\frac{R}{c}\right)}{4\pi c^2\rho_0 R_1}
            ,}
        \end{equation}
with $R = \frac{M(x-Mct)+R_1}{1-M^2}$ the distance between the emission and observation points and:
        \begin{equation}
            R_1=\sqrt{(x-Mct)^2+(y^2+z^2)(1-M^2)}.
        \end{equation}

\section{Expression of the radiation force in the far field}

\tikzmath{
	\angdip = 73;
}

Now that we have determined the field radiated by the translating dipolar source, the next step is to compute the self radiation force acting on it. The radiation force exerted on an object is by definition the time average of the stress integrated over the surface $S(t)$ of this object:
    \begin{equation}
        \left<
            \mathbf{F_{rad}}
        \right> = \left<
            \iint_{\surf(t)}\overline{\overline{\sigma}}\;\mathbf{n}\;\mathrm{d}S
        \right>,
    \end{equation}
    with $\overline{\overline{\sigma}}$ the stress tensor, $\mathbf{n}$ a unit vector normal to the integration surface, $\mathrm{d}S$ an infinitesimal surface and $\left<f\right>$ the time average of a time-dependant periodic function $f$ over its period. Such integration is difficult to perform directly since generally the surface $\surf(t)$ of the object is moving. In addition, this expression is not tractable in the punctual source approximation. Hence, this integral is classically turned into an integral over a steady spherical surface in the far field $\sinf$ (i.e. with a radius $\rinf$ much larger than all other characteristic length of the problem) encompassing the source  by using  the divergence and Reynold transport theorems (see e.g. the review by \cite{arfm_baudoin_2020}). Compared to the classical calculation, an additional difficulty is induced here by the translation of source. This problem was solved by \cite{jfm_roux_2022} by considering a spherical surface $\sinf$ centered on the source, and hence translating at the same velocity $\mathbf{\vit}$ as the source in $\refr$, and adapting the classical calculation to this translating surface. This leads to the following integral expression of the radiation force expressed as a function of the first order acoustic field:
\begin{equation}\label{eq:force}
\boxed{
	\mg\force\md=
	\mg\iint_{\sinf}{\cg\pg\dens\frac{v_1^2}{2}-\frac{1}{\dens\son^2}\frac{\pun^2}{2}\pd \ide  -\dens\vun\otimes\vun\cd}\mathbf{n} dS\md
	+\mg\iint_\sinf(\mathbf{\vit }\cdot\mathbf{n_\infty})\dun\vun dS\md.
}
\end{equation}
The first term corresponds to the classical integral expression of the radiation force on a steady surface at infinity, while the second term appears due to the translation of the surface $\sinf$.

\section{Expression of the velocity, pressure and density fields in the far field.}

The next step is to compute the velocity, pressure and density fields at first order in the far field to compute this integral.
        
    \subsection{Expression of the velocity, pressure and density fields}
The expression of the velocity field can be simply obtained from the definition of the acoustic displacement:
        \begin{equation}
            \mathbf{v_1}=\frac{\partial \mathbf{u_1}}{\partial t}.
            \label{eq:velocity}
        \end{equation}
Then, the introduction of the acoustic displacement in the continuity equation (\ref{eq:continuity}) leads to:
        \[
            \frac{\partial \rho_1}{\partial t}+\frac{\partial}{\partial t}\Big(\rho_0\nabla\cdot\mathbf{u_1}\Big)=0,
        \]
        which gives:
 \begin{equation}
          \rho_1 = \frac{p_1}{c^2} =   -\rho_0\nabla\cdot\mathbf{u_1}. \label{eq:pressure}
 \end{equation}
        \subsection{Far field approximation}
  
 To simplify the calculation of these expressions and of integral (\ref{eq:force}), we introduce the following change of variables:
      \[ 
                \begin{cases}
                    x^* &= x-Mct,\\
                    y^* &= y,\\
                    z^* &= z,\\
                    t^* &= t,\\
                \end{cases}
\qquad\text{and the associated spherical coordinates:}\qquad
                \begin{cases}
                    x^* &= r^*\cos\theta^*,\\
                    y^* &= r^*\sin\theta^*\cos\phi^*,\\
                    z^* &= r^*\sin\theta^*\sin\phi^*,\\
                \end{cases}
            \]
 corresponding to the Galilean transformation from $\mathcal{R}$ to $\mathcal{R}^* = (O^*,(x^* ,y^* ,z^* ),t^*)$ where $\mathcal{R}^*$ is the reference frame associated with the moving source.
\begin{figure}
	\centering
	\begin{tikzpicture}[scale=1]
\draw[->,color=BurntOrange] (0,0)--++(6,0) node [above] {$x$};
\draw[->,color=BurntOrange] (0,0)--++(0,3) node [right] {$y$};
\draw[->,color=BurntOrange] (0,0)--++(-2,-2) node [below right] {$z$};
\node[color=BurntOrange] at (-1,1.3) []{$\refr$};

\draw[->,color=TealBlue] (3,0)--++(4,0) node [above] {$x^*$};
\draw[->,color=TealBlue] (3,0)--++(0,3) node [right] {$y^*$};
\draw[->,color=TealBlue] (3,0)--++(-2,-2) node [below right] {$z^*$};
\node[color=TealBlue] at (6,2) []{$\refr^*$};

\coordinate (O) at (3,0) ;
\coordinate (A) at ($(O)+(\angdip:0.7)$) ;
\coordinate (B) at ($(O)-(\angdip:0.7)$) ;
\draw[color=gray, line width=2pt] (A)--(B) ;
\draw [color=white,ball color=gray,smooth] (A) circle (0.2);
\draw [color=white,ball color=gray,smooth] (B) circle (0.2);
\draw[->, line width=0.9] (A)--($(A)+(\angdip:0.7)$) node[above]{$\mathbf{c}$};

\draw [->,line width=1.2 pt,color=DarkOrchid] (3,0) --++ (1.5,0) node [below left] {$U$};

\draw [-] (3,0) circle (2.5);
\draw [-] (5.5,0) arc (0:-180:2.5 and 1) ;
\draw [dashed] (5.5,0) arc (0:180:2.5 and 1) ;
\draw [dashed] (3,2.5) arc (90:270:1.8 and 2.5) ;
\draw [-] (3,2.5) arc (90:-90:1.8 and 2.5) ;
\node[] at (5,-2) []{$\surf_\infty$};

\coordinate (EY) at ($(O)+(0,2)$);
\coordinate (EZ) at ($(O)+(224:1)$);
\coordinate (EM) at ($(EZ)+(0,2)$);
\coordinate (EP) at ($(EM)+(2.45,0)$);
\coordinate (EH) at ($(O)+(4,0)$);
\coordinate (EV) at ($(O)+(0,4)$);
\draw [->,color=TealBlue] (O)--(EP) node[above]{$P$};
\node[color=TealBlue] at ($(O)!.45!(EP)+(0.9,0)$ )[]{$r^*=\rinf$};
\draw [dashed,color=TealBlue] (EY)--(EM);
\draw [dashed,color=TealBlue] (EZ)--(EM);
\draw [dashed,color=TealBlue] (EM)--(EP);
\draw [dashed,color=TealBlue] (O)--(EM);
\draw pic["$\theta^*$", draw=TealBlue, ->, angle eccentricity=1.5, angle radius=0.6cm, text=TealBlue]{angle=EH--O--EP};
\draw pic["$\phi^*$", draw=TealBlue, ->, angle eccentricity=1.5, angle radius=0.6cm, text=TealBlue]{angle=EV--O--EM};
	\end{tikzpicture}
	\caption{Sketch illustrating the spherical coordinates $(r^*,\theta^*,\varphi^*)$ of a point $P$  in $\refr^*$.}
	\label{fig:schema_spherique} 
\end{figure}
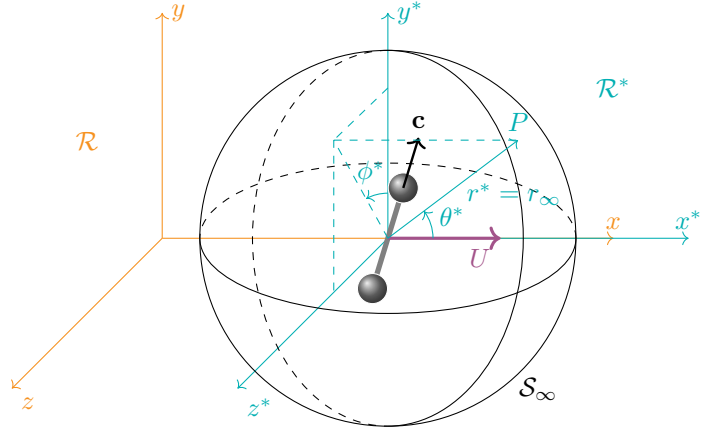

Assuming that the dipole velocity is small compared to the speed of sound ($M\ll1$) leads to the following expressions at first oder in $M$ (\cite{jfm_roux_2022}):
           \begin{subnumcases}{}
               R_1 = r^* & $+\mathscr{O}(M^2)$\\
                R = r^*(1+M\cos\theta^*) & $+\mathscr{O}(M^2)$
            \end{subnumcases}
            
            We can also compute the derivatives of these terms, which will be used later on :
            \begin{align*}
                \frac{\partial R  }{\partial t  } &\simeq -Mc(\cos\theta^*+M),    &
                \frac{\partial R_1}{\partial t  } &\simeq -Mc\cos\theta^*,        \\
                \frac{\partial R  }{\partial x^*} &\simeq \cos\theta^*+M,         &
                \frac{\partial R_1}{\partial x^*} &\simeq \cos\theta^*,           \\
                \frac{\partial R  }{\partial y^*} &\simeq \sin\theta^*\cos\phi^*, &
                \frac{\partial R_1}{\partial y^*} &\simeq \sin\theta^*\cos\phi^*, \\
                \frac{\partial R  }{\partial z^*} &\simeq \sin\theta^*\sin\phi^*, &
                \frac{\partial R_1}{\partial z^*} &\simeq \sin\theta^*\sin\phi^*.
            \end{align*}
            
        \subsubsection{Velocity field}
        
         Developping equation (\ref{eq:velocity}) within the small Mach number approximation leads to:
            \begin{eqnarray*}
                \mathbf{v_1} &=& \frac{1}{ 4\pi c^2\rho_0}\left[
                    \frac{1}{R_1}\frac{\partial \mathbf{f}\left(t-\frac{R}{c}\right)}{\partial t}
                    - \frac{1}{R_1^2}\frac{\partial R_1}{\partial t}\mathbf{f}\left(t-\frac{R}{c}\right)
                \right]\\
                &\simeq& \frac{1}{4\pi c^2\rho_0 r^*}\left[
                    \left(1-\frac{1}{c}\frac{\partial R}{\partial t}\right)\mathbf{f'}\left(t-\frac{R}{c}\right)
                    -\frac{1}{r^*}\frac{\partial R_1}{\partial t}\mathbf{f}\left(t-\frac{R}{c}\right)
                \right]\\
                &\simeq& \frac{1}{4\pi c^2\rho_0 r^*}\left[
                    (1+M\cos\theta^*)\mathbf{f'}\left(t-\frac{R}{c}\right)
                    +\frac{Mc\cos\theta^*}{r^*}\mathbf{f}\left(t-\frac{R}{c}\right)
                \right],
            \end{eqnarray*}
   which can be simplified in the far field approximation into:
            \begin{equation}
            \boxed{
                \mathbf{v_1} \simeq \frac{1+M\cos\theta^*}{4\pi c^2\rho_0 r^*}\mathbf{f'}\left(t-\frac{R}{c}\right).
                } \label{velocity}
            \end{equation}
            
        \subsubsection{Pressure field}
 Doing similar developments for the pressure field starting from Eq. (\ref{eq:pressure}) leads to:
            \begin{eqnarray*}
                p_1 &=& -\frac{1}{4\pi}\nabla\cdot\frac{\mathbf{f}\left(t-\frac{R}{c}\right)}{R_1}\\
                &\simeq&-\frac{1}{4\pi R_1^2}\left[
                    R_1\nabla\cdot\mathbf{f}\left(t-\frac{R}{c}\right) -
                    \mathbf{f}\left(t-\frac{R}{c}\right)\cdot\nabla R_1
                \right]\\
                &\simeq&-\frac{1}{4\pi}\left[
                    \frac{1}{r^*}\left(-\frac{1}{c}\right)\nabla R\cdot\mathbf{f'}\left(t-\frac{R}{c}\right)
                    -\frac{1}{r^{2}_{\infty}}\nabla R_1\cdot\mathbf{f}\left(t-\frac{R}{c}\right)
                \right]\\
                &\simeq&\frac{1}{4\pi c r^*}\nabla R\cdot\mathbf{f'}\left(t-\frac{R}{c}\right),
            \end{eqnarray*}
            i.e.
            \begin{equation}
            \boxed{
                p_1\simeq\frac{1}{4\pi c r^*}
                    \begin{pmatrix}
                        \cos\theta^*+M\\\sin\theta^*\cos\phi^*\\\sin\theta^*\sin\phi^*
                    \end{pmatrix}\cdot\mathbf{f'}\left(t-\frac{R}{c}\right),
                    }
                \label{pressure}
            \end{equation}
            in the far field approximation, at first order in $M$.
            \begin{figure}
                \centering
                    \includegraphics[width=\textwidth]{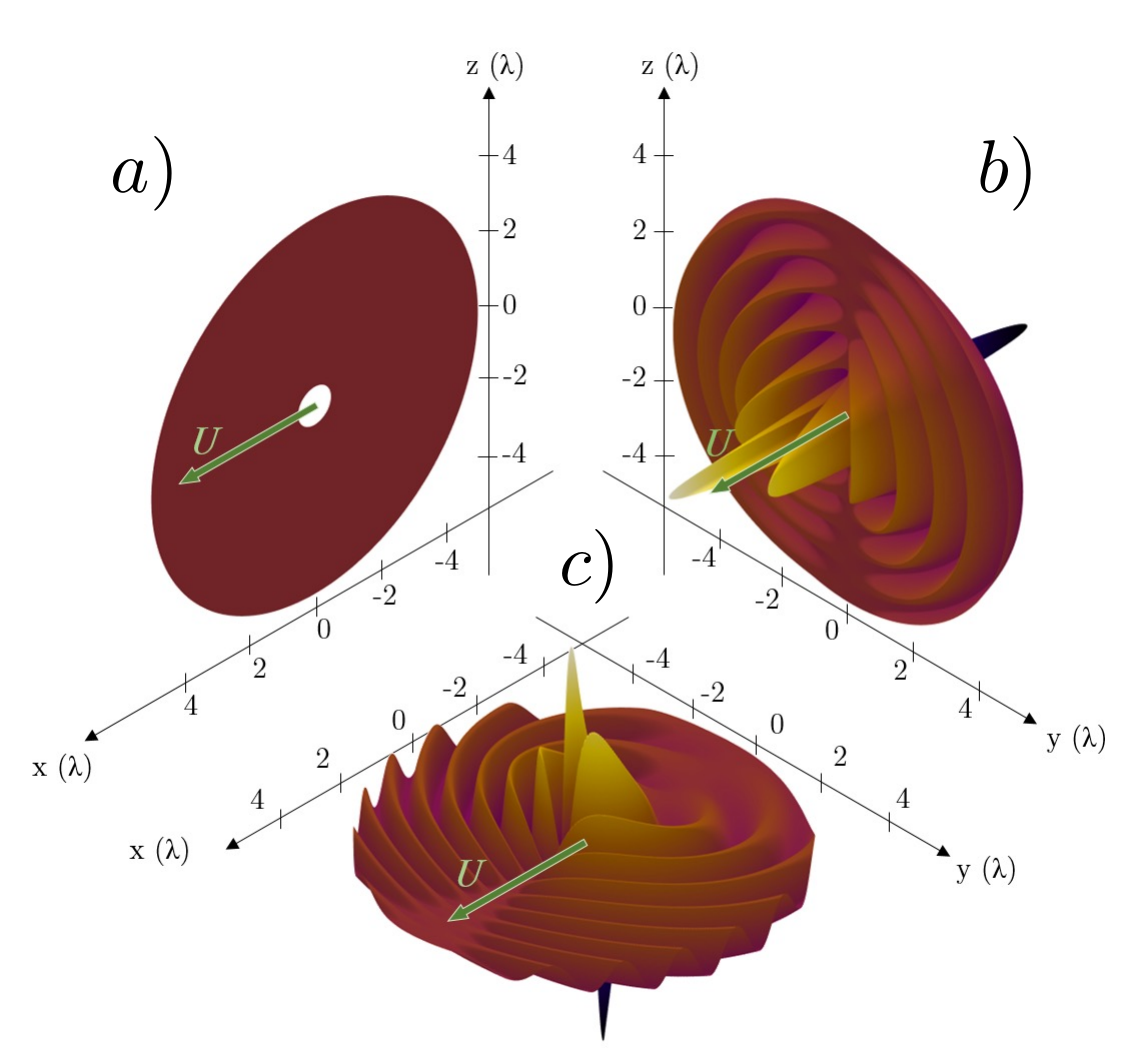}
                \caption{Normalized pressure field created by a dipolar source aligned along y-axis and translating along x-axis calculated with equation \eqref{pressure} represented in (a) the (O,x,z) plane, (b) the (O,y,z) plane and (c) the (O,x,y) plane. To magnify the asymmetry induced by Doppler effect, the pressure field is plotted for a Mach number $M = 0.5$.}
                \label{fig:pressure_field}
            \end{figure}
            This pressure field for a translating dipole is represented on figure \ref{fig:pressure_field} in the 3 orthogonal planes $(O, x, y)$, $(O, x, z)$ and $(O, y, z)$ for a dipole aligned with $y$ axis (i.e. for $c_x = c_z = 0$ and $c_y = 1$). In this representation, the Mach number was set to 0.5 to magnify the asymmetry induced by Doppler effect (visible on figure \ref{fig:pressure_field}.c).
            
                    \subsubsection{Density field}
            
 The density $\rho_1$ is simply equal to:
 \begin{equation}
 \boxed{
  \rho_1 = p_1 / c^2 \label{density}
  }
  \end{equation}          
        \subsection{Expression of the derivative of the dipolar source}
 
 We can see that the time derivative of the force $ \mathbf{f'}(t)$ appears in all these expressions, which according to equation (\ref{eq:f}) is equal to:
            \begin{equation}
                \mathbf{f'}(t) \quad=\quad\begin{pmatrix}
                    f'_{x}\\
                    f'_{y}\\
                    f'_{z}
                \end{pmatrix}\quad=\quad  F\omega \cos(\omega t)
                \begin{pmatrix}
                    c_{x}\\
                    c_{y}\\
                    c_{z}\\
                \end{pmatrix},
                \label{eq:f'}
            \end{equation}
  and whose mean square is equal to:
            \begin{equation}
                \left<\mathbf{f'}^2\right> = \frac{F^2\omega^2}{2}. \label{meansquare}
            \end{equation}
            
\section{Radiation force}

The last step is to compute integral (\ref{eq:force}) with the expression of the velocity, pressure and density fields given by equations (\ref{velocity}) to (\ref{density}). The radiation force $\mg \force \md$ can be decomposed into 4 contributions: the kinetic energy contribution $\mg \mathbf{F_{k}} \md$, the potential energy contribution $\mg \mathbf{F_{p}} \md$, the convective contribution $\mg \mathbf{F_c} \md$ and the translation contribution $\mg \mathbf{F_t} \md$:

    \begin{equation}
        \mg \mathbf{F_{rad}} \md =  \mg \mathbf{F_{k}}  \md  - \mg  \mathbf{F_{p}}  \md  -  \mg  \mathbf{F_c}   \md +  \mg \mathbf{F_t}  \md,
        \label{eq:radiation}
    \end{equation}
whose expressions are given by:
    \begin{subequations}
    \begin{align}
        \mg \mathbf{F_{k}} \md &= \left<\iint_{S_{\infty}}\frac{\rho_0}{2}\mathbf{v_1}^2\mathbf{n}_{\infty}\mathrm{d}S\right>\label{eq:rad_cin} \\
        \mg \mathbf{F_{p}} \md &= \left<\iint_{S_{\infty}}\frac{1}{\rho_0 c^2}\frac{p_1^2}{2}\mathbf{n}_{\infty}\mathrm{d}S\right>\label{eq:rad_pot }\\
        \mg \mathbf{F_c} \md &= \left<
                \iint_{S_{\infty}}\rho_0(\mathbf{v_1}\otimes\mathbf{v_1})\mathbf{n}_{\infty}\d S
            \right>\label{eq:rad_con}\\
        \mg \mathbf{F_t} \md &= \left<\iint_{S_{\infty}}(\mathbf{U}\cdot\mathbf{n}_{\infty})\rho_1\mathbf{v_1}\mathrm{d}S\right>\label{eq:rad_tra}
    \end{align}
    \end{subequations}
    
    \subsection{Potential energy term}
Let's start with the potential energy term:
        \begin{equation}
            \mg \mathbf{F_{p}} \md = \frac{1}{32\pi^2\rho_0 c^4}
                \int_{\theta^*=0}^{\pi}\int_{\phi^*=0}^{2\pi}
                    \left<
                        \left(\nabla R\cdot \mathbf{f'}\left(t-\frac{R}{c}\right)\right)^2
                        \sin\theta^*\mathbf{n}_{\infty}
                    \right>
                \d\theta^*\d\phi^*.
        \end{equation}
        
 We have:
        \begin{eqnarray*}
            \left(\nabla R\cdot \mathbf{f'}\left(t-\frac{R}{c}\right)\right)^2\sin\theta^*\mathbf{n}_{\infty}
            &=& \left(
                \begin{pmatrix}
                    \cos\theta^*+M\\\sin\theta^*\cos\phi^*\\\sin\theta^*\sin\phi^*
                \end{pmatrix}\cdot\begin{pmatrix}
                    f_{x}'\\f_{y}'\\f_{z}'
                \end{pmatrix}
            \right)^2\begin{pmatrix}
                \cos\theta^*\sin\theta^*\\\sin^2\theta^*\cos\phi^*\\\sin^2\theta^*\sin\phi^*
            \end{pmatrix},
        \end{eqnarray*}
with
        \begin{align}
            \left(
                \begin{pmatrix}
                    \cos\theta^*+M\\\sin\theta^*\cos\phi^*\\\sin\theta^*\sin\phi^*
                \end{pmatrix}\cdot\begin{pmatrix}
                    f_{x}'\\f_{y}'\\f_{z}'
                \end{pmatrix}
            \right)^2
            & =(\cos\theta^*+M)^2 f_{x}'^2 + \sin^2\theta^*\cos^2\phi^*f_{y}'^2+\sin^2\theta^*\sin^2\phi^*f_{z}'^2 \nonumber \\
            & +2(\cos\theta^*+M)\sin\theta^*\cos\phi^*f_{x}'f_{y}' \nonumber \\
            & +2(\cos\theta^*+M)\sin\theta^*\sin\phi^*f_{x}'f_{z}' \nonumber \\
            & +2\sin\theta^*\sin\theta^*\cos\phi^*\sin\phi^*f_{y}'f_{z}'. \nonumber
        \end{align}

Let's take a closer look at $<f_{i}'f_{j}'>$, with $i$ and $j$ two indexes taken among $\{x,y,z\}$. From equation (\ref{eq:f'}), we can write:
        \begin{eqnarray*}
            <f_{i}'f_{j}'> &=& F^2\omega^2\left<\cos^2\left(\omega\left(t-\frac{R}{c}\right)\right)\right> c_i c_j\\
            &=& \frac{F^2\omega^2}{2}
                \left<
                    \cos(\psi_i-\psi_j)
                    +\cos\left(
                        2\omega t
                        -2\omega\frac{1+M\cos\theta^*}{c}r_{\infty}
                        +\phi_i+\phi_j
                    \right)
                \right>c_ic_j\\
            &=&\frac{F^2\omega^2}{2}c_ic_j.
        \end{eqnarray*}
        Hence, within the approximation of small velocity perturbation, we see that $<f_{i}'f_{j}'>$ does not depend on $\theta^*$; we can therefore take these terms out of the integral. Then, the terms in $\cos\phi^*$, $\sin\phi^*$, and $\cos\phi^*\sin\phi^*$ cancel when they are integrated over $[0,2\pi]$, as well as the ones in $\cos^3\theta^*\sin\theta^*$, $\cos\theta^*\sin^3\theta^*$, $\cos\theta^*\sin\theta^*$, and $\cos^3\theta^*\sin^3\theta^*$ when they are integrated over $[0,\pi]$. These simplifications lead to:
        \begin{equation}
        \boxed{
            \mg \mathbf{F_{p}} \md=\frac{MF^2\omega^2}{24\pi\rho_0 c^4}
                \begin{pmatrix}
                    c_{x}^2\\
                    c_{x}c_{y}\\
                    c_{x}c_{z}
                \end{pmatrix}.
                }
            \label{eq:potentialterm}
        \end{equation}
        
    \subsection{Kinetic energy term}
Now let's focus on the kinetic energy term:
        \begin{equation}
            \mg \mathbf{F_{k}} \md = \int_{\theta^*=0}^{\pi}\int_{\phi^*=0}^{2\pi}\left<\frac{\rho_0}{2}\mathbf{v_1}^2\mathbf{n}_{\infty}r_{\infty}^2\sin\theta^*\right>\d\phi^*\d\theta^*.
        \end{equation}
   We have:
        \begin{equation}
            \left<\frac{\rho_0}{2}\mathbf{v_1}^2\mathbf{n}_{\infty}r_{\infty}^2\sin\theta^*\right>
            = \frac{(1+M\cos\theta^*)^2}{32\pi^2c^4\rho_0}\left<\mathbf{f'}^2\right>\cdot\begin{pmatrix}
                \cos\theta^*\sin\theta^*\\\sin^2\theta^*\cos\phi^*\\\sin^2\theta^*\sin\phi^*
            \end{pmatrix},\\
        \end{equation}
which with the help of equation (\ref{meansquare}) turns into: 
        \begin{equation}
            \left<\frac{\rho_0}{2}\mathbf{v_1}^2\mathbf{n}_{\infty}r_{\infty}^2\sin\theta^*\right>
            =\frac{(1+M\cos\theta^*)^2}{32\pi^2c^4\rho_0}\frac{F^2\omega^2}{2}\cdot\begin{pmatrix}
                \cos\theta^*\sin\theta^*\\\sin^2\theta^*\cos\phi^*\\\sin^2\theta^*\sin\phi^*
            \end{pmatrix}.
        \end{equation}
After integration over $\phi*$ and $\theta^*$, we get :
        \begin{equation}
        \boxed{
            \mathbf{F_{k}} = \frac{MF^2\omega^2}{24\pi c^4\rho_0} \mathbf{x}.
            \label{fk}
            }
        \end{equation}
        
    \subsection{Convective term}
        We then study:
        \begin{equation*}
            \mg \mathbf{F_c} \md = \left<
                \iint_{S_{\infty}}\rho_0(\mathbf{v_1}\otimes\mathbf{v_1})\mathbf{n}_{\infty}\d S
            \right>.
        \end{equation*}
        We have :
        \[
            \rho_0r_{\infty}^2\sin\theta^*(\mathbf{v_1}\otimes\mathbf{v_1})\mathbf{n}_{\infty} = \frac{(1+M\cos\theta^*)^2}{16\pi^2 c^4\rho_0}
            \begin{pmatrix}
                f_{x}^{\prime2} & f'_{x} f'_{y} & f'_{x} f'_{z}  \\
                f'_{y} f'_{x} & f_{y}^{\prime2} & f'_{y} f'_{z}\\
                f'_{z} f'_{x} & f'_{z} f'_{y} & f_{z}^{\prime2}
            \end{pmatrix}\begin{pmatrix}
                \cos\theta^*\sin\theta^*\\
                \sin^2\theta^*\cos\phi^*\\
                \sin^2\theta^*\sin\phi^*
            \end{pmatrix}.
        \]

        When integrating over $\phi^*$, we find that the terms in $\cos\phi^*$ and $\sin\phi^*$ cancel so that:
        \begin{eqnarray*}
            \mg \mathbf{F_c} \md &=&
                2\pi\int_0^{\pi}\left<
                    \frac{(1+M\cos\theta^*)^2}{16\pi^2 c^4\rho_0}
                    \cos\theta^*\sin\theta^*
                    \begin{pmatrix}
                        f_{x}^{\prime2}\\
                        f'_{y} f'_{x}\\
                        f'_{z} f'_{x}
                    \end{pmatrix}\right>
                \d\theta^*\\
            &=& \frac{F^2\omega^2}{16\pi c^4\rho_0}
                \begin{pmatrix}
                    c_x^2\\
                    c_{y} c_{x}\\
                    c_{z} c_{x}
                \end{pmatrix}
                \int_0^{\pi}
                    (1+2M\cos\theta^*)\cos\theta^*\sin\theta^*
                \d\theta^*\\
            &=& \frac{MF^2\omega^2}{16\pi c^4\rho_0}
                \begin{pmatrix}
                    c_x^2\\
                    \alpha_{yx} c_{y} c_{x}\\
                    \alpha_{zx} c_{z} c_{x}
                \end{pmatrix} \times \frac{4}{3}.
        \end{eqnarray*}
        Finally, we obtain:
        \begin{equation}
        \boxed{
            \mathbf{F_{c}}=\frac{MF^2\omega^2}{12\pi c^4\rho_0}
                \begin{pmatrix}
                    c_x^2\\
                   c_{y} c_{x}\\
                   c_{z} c_{x}
                \end{pmatrix}.
                }
        \end{equation}
        
    \subsection{Source translation term}
  And finally we can compute the translation term:
        \begin{equation*}
            \mg \mathbf{F_t} \md = \left<\iint_{S_{\infty}}(\mathbf{U}\cdot\mathbf{n}_{\infty})\rho_1\mathbf{v_1}\mathrm{d}S\right>.
        \end{equation*}
We have:
        \begin{eqnarray*}
            (\mathbf{U}\cdot\mathbf{n}_{\infty})\rho_1\mathbf{v_1}r_{\infty}^2\sin\theta^*
            &=& Mc \; \mathbf{x} \cdot
                \begin{pmatrix}
                    \cos\theta^*\sin\theta^*\\
                    \sin^2\theta^*\cos\phi^*\\
                    \sin^2\theta^*\sin\phi^*
                \end{pmatrix}\frac{p_1}{c^2}\mathbf{v_1}r_{\infty}^2\\
            &=&M\cos\theta^*\sin\theta^*
                \frac{1+M\cos\theta^*}{16\pi^2c^4\rho_0}
               \left[  \begin{pmatrix}
                    \cos\theta^*+M\\\sin\theta^*\cos\phi^*\\\sin\theta^*\sin\phi^*
                \end{pmatrix}\cdot
                \begin{pmatrix}
                    f'_{x}\\
                    f'_{y}\\
                    f'_{z}
                \end{pmatrix} \right]
                \begin{pmatrix}
                    f'_{x}\\
                    f'_{y}\\
                    f'_{z}
                \end{pmatrix}.
        \end{eqnarray*}
        As previously, the terms in $\sin\phi^*$ and $\cos\phi^*$ cancel when integrating, leading to:
        \begin{eqnarray*}
            \mg \mathbf{F_t} \md &=& \int_0^{\pi}
                M\cos\theta^*\sin\theta^*
                \frac{1+M\cos\theta^*}{8\pi c^4\rho_0}
                (\cos\theta^*+M)\left<
                \begin{pmatrix}
                    f'_xf'_x\\
                    f'_xf'_y\\
                    f'_xf'_z
                \end{pmatrix}\right>
            \d\theta^*\\
            &=& \frac{MF^2\omega^2}{16\pi c^4\rho_0}
                \int_0^{\pi}
                    \cos\theta^*\sin\theta^*
                    (1+M\cos\theta^*)
                    (\cos\theta^*+M)
                    \begin{pmatrix}
                        c_x^2\\
                        c_xc_y\\
                        c_xc_z
                    \end{pmatrix}
                \d\theta^*\\
            &=&\frac{MF^2\omega^2}{16\pi c^4\rho_0}
            \begin{pmatrix}
                c_x^2\\
                c_xc_y\\
                c_xc_z
            \end{pmatrix} \times \frac{2}{3}.
        \end{eqnarray*}
We finally obtain:
        \begin{equation}
        \boxed{
            \mg \mathbf{F_t} \md = \frac{MF^2\omega^2}{24\pi c^4\rho_0}
                \begin{pmatrix}
                    c_x^2\\
                    c_xc_y\\
                    c_xc_z
                \end{pmatrix}.
                }
                \label{ft}
        \end{equation}
    
    \subsection{Total radiation force}
If we now gather equations (\ref{fk}) to (\ref{ft}), we obtain:
        \begin{eqnarray*}
        \mg \mathbf{F_{rad}} \md & = & \mg \mathbf{F_{k}}  \md  - \mg  \mathbf{F_{p}}  \md  -  \mg  \mathbf{F_c}   \md +  \mg \mathbf{F_t}  \md \\
            &=& \frac{MF^2\omega^2}{24\pi c^4\rho_0} \mathbf{x}
            +\begin{pmatrix}
                    c_{x}^2\\
                    c_{x}c_{y}\\
                    c_{x}c_{z}
                \end{pmatrix}
            \left(
            -\frac{MF^2\omega^2}{24\pi\rho_0 c^4}
            - \frac{MF^2\omega^2}{12\pi c^4\rho_0}
            + \frac{MF^2\omega^2}{24\pi c^4\rho_0}
            \right),     
        \end{eqnarray*}
   and thus the final expression of the self-induced radiation force acting on a translating dipolar source:
        \begin{equation}
        \boxed{
            \mg \mathbf{F_{rad}} \md = \frac{MF^2\omega^2}{12\pi c^4\rho_0}
                \begin{pmatrix}
                    1/2 - c_{x}^2\\
                    -c_{x}c_{y}\\
                    -c_{x}c_{z}
                \end{pmatrix}.}
            \label{fradf}
        \end{equation}
        
\section{Discussion}

As expected, this expression shows that the radiation force only exists in presence of a velocity perturbation i.e. for $U = M c \neq 0$ and the radiation force is proportional to the intensity of the wave radiated by the source. An interesting point is that the projection of this radiation force along the velocity perturbation axis $\mathbf{x}$ can be either positive or negative depending on the orientation of the dipole. To further analyse this expression, it is interesting to introduce the spherical coordinates ($\hat{\theta}$, $\hat{\phi}$) of the unit vector $\mathbf{c}$ which sets the direction of the dipolar radiation:
        \begin{subnumcases}{}
c_x = \cos \hat{\theta}, \\
c_y =  \sin \hat{\theta} \cos \hat{\phi}, \\
c_z = \sin \hat{\theta} \sin \hat{\phi}.
        \end{subnumcases}
If we replace these expressions in (\ref{fradf}), and compute the norm of the radiation force, we obtain:
\begin{eqnarray}
\| \mg \force \md \| & = & \frac{MF^2\omega^2}{12\pi c^4\rho_0} \sqrt{\left( \frac{1}{2} - \cos^2 \hat{\theta}  \right)^2 + \cos^2 \hat{\theta}   \sin^2 \hat{\theta}  \cos^2 \hat{\phi} +   \cos^2 \hat{\theta}   \sin^2 \hat{\theta}  \sin^2 \hat{\phi}  } \nonumber \\
& = &  \frac{MF^2\omega^2}{12\pi c^4\rho_0} \sqrt{  \frac{1}{4} -  \cos^2 \hat{\theta} + \cos^4 \hat{\theta}  + \cos^2 \hat{\theta} \sin^2 \hat{\theta} } \nonumber \\
& = &  \frac{MF^2\omega^2}{12\pi c^4\rho_0} \sqrt{  \frac{1}{4} + \cos^2\hat\theta \left( -1 +  \cos^2 \hat{\theta} +  \sin^2 \hat{\theta} \right) } \nonumber \\
& = & \frac{MF^2\omega^2}{24\pi c^4\rho_0}.
\end{eqnarray}

This calculation shows that the radiation force magnitude is independent of the angles ($\hat{\theta}$, $\hat{\phi}$) between the dipole orientation and the perturbation displacement. 

Let's now consider the above calculation as a stability analysis, with a stable state corresponding to the dipolar source at rest perturbed by a small velocity perturbation. Then the most unstable situation arises when the projection of the force along the velocity perturbation $\mathbf{x}$ is maximum, since in this case the amplification of the velocity perturbation by the radiation force will be maximum.
 Since the magnitude of the radiation force is constant whatever the orientation of the dipole, this happens when $\mg \force \md$ is directed in the $\mathbf{x}$ direction, i.e. when $\hat{\theta} = \frac{\pi}{2} $. The same result is of course obtained if we maximize $\mg \force \md . \mathbf{x} = \frac{1}{2} - \cos^2 \hat{\theta}$.  Since the most unstable situations correspond to a dipole direction orthogonal displacement, an interesting point is that the direction of motion is not set a priori by the anisotropy of the source as in recent experiments with capillary and gravity waves (\cite{arxiv_ho_2021,prf_benham_2022}) surfers.

\section{Conclusion and perspectives}

In this paper, we compute the self-radiation force exerted on a dipolar source in presence of a small axial velocity perturbation. We show that when the dipole orientation is orthogonal to the velocity perturbation axis, the radiation force is aligned with the motion hence amplifying it. This result suggests the possibility of an acoustical source surfing on its own acoustic field, a prerequisite for pilot waves analogues. From a theoretical point of view, a next step would be to derive constitutive equations for these acoustic surfers dynamics. It would also be interesting to investigate the response of an arbitrary multipole source since our result with monopolar and dipolar sources exhibit drastically different behaviours. Of course the most compelling challenge would be to materialize these acoustic surfers experimentally. Many different types of acoustic sources and surrounding fluids can be envisionned. A first requirement is to find a way to excite these acoustic surfers in absence of any incident directional acoustic field. Otherwise the source will be driven by this incident field and not by its own wave. Different possibilities can be considered as (i) using a non-acoustic homogeneous field to excite the source (acceleration, magnetic, electric) or (ii) creating a stochastic acoustic field which does not prescribe the motion in any specific direction. A second requirement is that the radiation force should overcome the drag force for the situation to be unstable. This might require to work in low viscosity fluids, such as gases, cryogenic fluids or superfluids.  

This work hence paves the way for further theoretical, numerical and experimental investigation of acoustic quantum analogues.

\section*{Acknowledgements}
We acknowledge support from ISITE ERC Generator program and stimulating discussions with Pr. J. Bush, which motivated us to perform this work.

\section*{Declaration of Interests}

The authors report no conflict of interest.


\end{document}